\documentclass{ws-ijmpa}

\begin{document}

\markboth{A. Valcarce, J. Vijande}
{On the nature of the X(3872)}

%
\catchline{}{}{}{}{}
%

\title{ON THE NATURE OF THE X(3872)}

\author{A. VALCARCE}

\address{Departamento de F\'\i sica Fundamental\\ Universidad de Salamanca, Salamanca, Spain\\
valcarce@usal.es}

\author{J. VIJANDE}

\address{Departamento de F\'\i sica At\'omica, Molecular y Nuclear\\ Universidad de Valencia (UV)
and IFIC (UV-CSIC), Valencia, Spain\\
javier.vijande@uv.es}

\maketitle

\begin{history}
\received{Day Month Year}
\revised{Day Month Year}
\end{history}

\begin{abstract}
We present recent studies of charmonium multiquark states.
We use different interacting models and numerical methods to study deeply bound 
four-quark states and meson-meson molecules. No deeply bound four-quark
states are found in our analysis. A nice description of 
the X(3872) is obtained as a $D\overline{D}^*-J/\Psi\omega$
coupled channel state.
\end{abstract}

\ccode{PACS numbers: 14.40.Gx, 21.30.Fe, 12.39.Mk}

Since 2003 several states have been discovered in the charmonium mass region.
While in the conventional description
of charmonium in terms of quark-antiquark pairs
some states are still missing, the number of experimental states
reported up to now is larger than empty spaces in the
$c\bar c$ spectrum. This overpopulation, together with other difficulties
to explain observed states as simple quark-antiquark
pairs, triggered discussions~\cite{Bug10} on a possible exotic interpretation,
four-quark states either as compact tetraquarks or slightly
bound meson-meson molecules.

Understanding of charmonium spectroscopy is challenging
for experimentalists and theorists alike.
Charmonium has been used as the test bed to demonstrate the color
Fermi-Breit structure of quark atoms obeying the same principles
as ordinary atoms~\cite{Isg83}. Its nonrelativistic character
gave rise to an amazing agreement between
experiment and simple quark potential model predictions
as $c\bar c$ states~\cite{Eic80}.
The opening of charmed meson thresholds was expected to modify
the trend in the construction of quark-antiquark models.
In the adiabatic approximation
meson loops were absorbed into the static interquark potential.
Thus, close to the threshold production of charmed mesons
models required of an improved interaction~\cite{Isg99}.

The discussion above suggests that charmonium spectroscopy could be rather
simple below the threshold production of charmed mesons but
much more complex above it. In particular, the coupling to
the closest $(c\bar c)(n\bar n)$ system, referred to
as {\it unquenching the naive quark model}~\cite{Clo05},
could be an important spectroscopic ingredient. Therefore,
hidden-charm four-quark states could
explain the overpopulation of quark-antiquark theoretical states.
Thus, the new experimental discoveries are offering exciting
new insights into the subtleties of the strong interaction.

In an attempt to disentangle the role played by multiquark configurations
in the charmonium spectroscopy we obtained an exact solution of the
four-body problem based on an infinite expansion of the four-quark
wave function in terms of hyperspherical harmonics~\cite{Vij07}.
From our analysis, we concluded that those four-quark states with two
different asymptotic physical thresholds (as it is
the case of the $c\bar c n \bar n$ system that may split either
into a $(c\bar c)(n \bar n)$ or
$(c \bar n)(n \bar c)$ two-meson states) can hardly present a bound state
since the interaction between any pair of quarks contributes to the energy
of one of the two physical thresholds.
Close to a threshold we have used a different technique that we developed when studying
baryon spectra with screened potentials. We solved the Lippmann-Schwinger equation
looking for attractive channels that may contain a meson-meson molecule~\cite{Ter09}.
In order to account for all basis states we allow
for the coupling to charmonium-light two-meson systems.
Thus, we consider the system of two mesons $M_1$ and $\overline{M}_2$ ($M_i=D, D^*$)
in a relative $S-$state interacting through a potential $V$ that contains a
tensor force, and therefore there is a coupling to the
$M_1\overline{M}_2$ $D-$wave. Moreover,
the two $D$-meson system can couple to a charmonium-light
two-meson state, for example $D\overline{D}^*$ is coupled to
$J/\Psi \omega$.
We have consistently used the same interacting Hamiltonian to study
the two- and four-quark systems to guarantee that thresholds
and possible bound states are eigenstates of the same Hamiltonian.

\begin{table}[ph]
\tbl{Attractive channels for the two $D-$meson systems.}
{\begin{tabular}{@{}cccccc@{}} \toprule
System & $D\overline{D}$ & $D\overline{D}^*$ &\multicolumn{3}{c}{ $D^*\overline{D}^*$} \\ \colrule 
$J^{PC}(I)$ &  $0^{++}(0)$ & $1^{++}(0)$ & $0^{++}(0)$ & $2^{++}(0)$ & $2^{++}(1)$ \\ \botrule
\end{tabular} \label{t1}}
\end{table}

Our results are shown in Table~\ref{t1}.
This study has been the first systematic analysis of four-quark
hidden-charm states as compact states or meson-meson molecules. For the first
time we have performed a consistent study of all quantum numbers within the
same model.
Our predictions robustly show that no deeply bound states can be expected for this system.
Only a few channels can be expected to
present observable resonances or slightly bound states. Among them, we
have found that the $D\overline{D}^*$ system must show a bound state
slightly below the threshold for charmed mesons production
with quantum numbers $J^{PC}(I)=1^{++}(0)$,
that could correspond to the widely discussed $X(3872)$. Of the systems
made of a particle and its corresponding antiparticle,
$D\overline{D}$ and $D^*\overline{D}^*$, the $J^{PC}(I)=0^{++}(0)$
is attractive. It would be the only candidate to accommodate
a wide resonance for the $D\overline{D}$ system.
For the $D^*\overline{D}^*$ the attraction
is stronger and structures may be observed close and above the charmed
meson production threshold.
Also, we have shown that the $J^{PC}(I)=2^{++}(0,1)$ $D^*\overline{D}^*$
channels are attractive due to the coupling to the $J/\Psi \omega$ and
$J/\Psi \rho$ channels, respectively.

Among these exotic theoretical states, charged states have an
unique feature: by construction they cannot be accommodated into
the conventional $c\bar c$ spectrum. Two different experimental
findings show positive results on charge charmonium mesons.
The first one was a $\Psi(2S)\pi^+$ peak at about 
4430 MeV/c$^2$ observed by Belle in the
$\bar B^0 \to \Psi (2S)\pi^+K^-$ decays~\cite{Cho08}. A second positive
observation was reported by Belle from the $B^0 \to \Xi_{c1} \pi^+K^-$ decay,
with two resonances in $\Xi_{c1} \pi^+$ at masses of about
4050 and 4250 MeV/c$^2$~\cite{Miz08}.
In the first case, BaBar has presented its own analysis~\cite{Aub08}
performing a detailed study of the acceptance and possible
reflections concluding that no significant signal exists on
the data. While the two experiments made different conclusion,
the data itself seem to be in a reasonable agreement except
for the lower available statistics of the BaBar experiment.
The states found in Ref.~\cite{Miz08} could correspond to the
$D^*\overline{D}^*$ $J^{PC}(I)=2^{++}(1)$ we 
have predicted~\cite{Ter09}.
Its confirmation would represent a unique tool in discriminating
among different theoretical models.

Due to heavy quark symmetry, replacing the charm
quarks by bottom quarks decreases the kinetic energy without significantly
changing the potential energy. In consequence, four-quark bottomonium mesons
must also exist and have larger binding energies. An experimental effort
in this direction will confirm or rule out the theoretical expectations.
If the scenario presented here turns out to be correct, it will open a new
interesting spectroscopic area.

\section*{Acknowledgments}
This work has been partially funded by the Spanish Ministerio de
Educaci\'on y Ciencia and EU FEDER under Contract No. FPA2007-65748,
by Junta de Castilla y Le\'{o}n under Contract No. GR12,
and by the Spanish Consolider-Ingenio 2010 Program CPAN (CSD2007-00042).


\begin{thebibliography}{0} 

\bibitem{Bug10} D. V. Bugg, {\it J. Phys. G} 
			{\bf 37}, 055002 (2010).

\bibitem{Isg83} N. Isgur and G. Karl,
                        {\it Phys. Tod.} {\bf Nov. 1983}, 36 (1983).

\bibitem{Eic80} E. Eichten, K. Gottfried, T. Kinoshita, K. D. Lane and T.-M. Yan,
                        {\it Phys. Rev. D} {\bf 21}, 203 (1980).

\bibitem{Isg99} N. Isgur,
                        {\it Phys. Rev. D} {\bf 60}, 054013 (1999);
E. J. Eichten, K. Lane and C. Quigg,
                        {\it Phys. Rev. D} {\bf 69}, 094019 (2004).

\bibitem{Clo05} F. E. Close,
                        {\it arXiv:0706.2709}.

\bibitem{Vij07} J. Vijande, E. Weissman, N. Barnea and A. Valcarce,
                        {\it Phys. Rev. D} {\bf 76}, 094022 (2007).

\bibitem{Ter09} T. Fern\'andez-Caram\'es, A. Valcarce and J. Vijande,
			{\it Phys. Rev. Lett.} {\bf 103}, 222001 (2009).

\bibitem{Cho08} S. K. Choi {\it et al.} (Belle Collaboration),
                        {\it Phys. Rev. Lett.} {\bf 100}, 142001 (2008).

\bibitem{Miz08} R. Mizuk {\it et al.} (Belle Collaboration),
                        {\it Phys. Rev. D} {\bf 78}, 072004 (2008).

\bibitem{Aub08} B. Aubert {\it et al.} (BaBar Collaboration),
                        {\it Phys. Rev. D} {\bf 79}, 112001 (2009).
\end{thebibliography}
\end{document}